\documentclass[12pt]{revtex4}

\usepackage{graphicx,epsfig}
\usepackage{bm,longtable,enumerate}
\usepackage{amsmath,amssymb}

\def\beq{\begin{equation}}
\def\eeq{\end{equation}}
\def\beqa{\begin{eqnarray}}
\def\eeqa{\end{eqnarray}}

\def\R{\rm l\!R\,}

\begin{document}

\title{ A conical deficit in the $AdS_4/CFT_3$ correspondence}

\author{C.~A.~Ballon Bayona $^{1}$}
\email{
 ballon@cbpf.br}

\author{Cristine N. Ferreira $^{2}$} 

\email{crisnfer@pq.cnpq.br}

\author{V.~J.~Vasquez~Otoya $^{1}$}

\email{vjose@cbpf.br}

\affiliation{\\
$^{1}$ Centro Brasileiro de Pesquisas F\'{i}sicas, \\ 
Rua Dr. Xavier Sigaud 150, Urca, 22290-180, Rio de Janeiro, RJ, Brazil \\
$^{2}$ N\'ucleo de Estudos em F\'{\i}sica, Instituto Federal de  Educa\c{c}\~ao, \\ Ci\^encia e Tecnol\'ogia Fluminense
28030-130 Campos, RJ, Brazil}

\date{\today}

\begin{abstract}
Inspired by the AdS/CFT correspondence we propose a new duality that allow the study of strongly coupled field theories living in a $2+1$ conical space-time.  Solving the  4-d Einstein equations in the presence of an infinite static string and negative cosmological constant we obtain a conical $AdS_4$ space-time whose boundary is identified with the  $2+1$ cone found by Deser, Jackiw and 't Hooft.  Using the $AdS_4/CFT_3$ correspondence we calculate retarded Green's functions of scalar operators living in the cone. 
\end{abstract}

\pacs{}

\maketitle

\section{Introduction}
The $AdS_{d+1}/CFT_{d}$ correspondence, discovered by J. Maldacena \cite{Maldacena:1997re}, has motivated in recent years the study of conformal gauge theories arising from 10-d string theory or 11-d M-theory. Most of the effort has been concentrated on the $d=4$ case where a configuration of $N$ coincident D3-branes leads to a duality between type IIB string theory in $AdS_5 \times S^5$ and ${\cal N}=4$ $SU(N)$ Super Yang-Mills Theory in $\R^{1,3}$ (some reviews can be found in \cite{Aharony:1999ti,D'Hoker:2002aw,Nastase:2007kj}). The $AdS_5/CFT_4$ correspondence brought also new approaches to the strong coupling problem in  QCD  (see for instance \cite{Erdmenger:2007cm,Gubser:2009md}). 

Recently, the $d=3$ case has gained a lot of attention due to the following two reasons. The first reason is the important progress made in the understanding of $D2$ and $M2$-branes configurations. It has been conjectured that M-theory in $AdS_4 \times S^7/ {\mathbb{Z}}_k$ is the dual of ${\cal N}=6$ $U(N)\times U(N)$  Super Chern Simons gauge theory with level $k$ living in $2+1$ Minkowski space-time \cite{Aharony:2008ug} (for a recent review see \cite{Klebanov:2009sg}). The second reason is the possible application of the $AdS_4/CFT_3$ correspondence to the phenomenology of $2+1$ strongly coupled systems in condensed matter \cite{Herzog:2009xv,Hartnoll:2009sz}. An interesting example is the graphene (a monolayer of carbon cristal) whose electrical conductivy has been recently predicted \cite{Hartnoll:2009sz}.

The $AdS_4/CFT_3$ correspondence allow the study of $2+1$ systems assuming a flat space-time. However, in real experiments 
(like that of graphene)  a $2+1$ system may present structural defects \cite{ReviewGraphene}. These kind of defects can be modeled by considering field theories in curved spaces like spheres and cones and may also modify the electronic and transport properties of the system (see for instance \cite{Lammert,Cortijo:2006xs}).

In this paper we propose, motivated in the $AdS_4/CFT_3$ correspondence,  a new duality that allow the study of strongly coupled field theories living in a conical space-time. First we find a conical $AdS_4$ space-time by solving the $3+1$ Einstein equations in the presence of an infinite static string and a negative cosmological constant.  We identify the boundary of this space-time with the  conical space-time of  Deser, Jackiw and 't Hooft obtained as a solution of the  $2+1$ Einstein equations in the presence of a massive point particle and zero cosmological constant \cite{Deser:1983tn}. In order to investigate the effect of the conical deficit on a  $2+1$ strongly coupled field theory, we calculate retarded Green's functions of  scalar operators ${\cal O}$.  Assuming a conical $AdS_4/CFT_3$ correspondence, a scalar operator ${\cal O}$ couples to  a scalar field $\phi$ living in the conical $AdS_4$ space-time so that real time retarded Green's functions of the boundary scalar operator ${\cal O}$  can be obtained from the on-shell action of the bulk scalar field $\phi$ using a prescription similar to that found by Son and Starinets \cite{Son:2002sd}. 

The conical $AdS_4$ space-time found in this paper can be compared in the limit of weak gravitational coupling to that obtained in \cite{Dehghani:2001ft} when considering a vortex line configuration. Black hole solutions in $AdS_4$ may also contain conical deficits \cite{Emparan:1999wa,Dehghani:2001nz}.

\section{A conical deficit in 4-d AdS space-time}

\subsection{The static string configuration}

Consider an  infinite static string living in a 4-d space-time with metric $g_{\mu \nu}$ in the presence of a negative cosmological constant $\Lambda=-3/L^2$. This system is described by the Einstein-Hilbert Nambu-Goto action 
\beqa
S &\,=\,& S_G \,+\, S_{NG} \nonumber \\
&\,=\,&  \frac{1}{16 \pi G_4} \int dt dz  dr d \theta \sqrt{-g} \left[ R + \frac{6}{L^2} \right ] - \mu \int d \sigma^0 d \sigma^1 \sqrt{ - \det{ P [g_{ab}]}} \, , \label{gravityplusstring}
\eeqa
where  
\beq
P [g_{ab}]\,=\, g_{\mu\nu}(X) \, \frac{\partial X^{\mu}(\sigma)}{\partial \sigma^a} \frac{\partial X^{\nu}(\sigma)}{\partial \sigma^b}  \, , 
\eeq
is the 2-d induced metric of the Nambu-Goto string with $g_{\mu \nu}(X)$ the 4-d space-time metric. Note that the embedding string coordinates $X^{\mu}(\sigma)$ are in general different from the space-time coordinates $\{ t, z, r , \theta \}$. Now we consider the configuration of a  static string extended in the $z$ direction. The appropiate gauge for this scenario is $ \{\sigma^0,\sigma^1 \} =  \{t, z \}$ and the embedding string coordinates $X^{\mu}(t, z) = \{t, z,X^1(z),X^2(z) \}$. 

Varying the Nambu-Goto action with respect to $X^1(z)$ and $X^2(z)$ we obtain the equations of motion
\beqa
\partial_z \Big [  \frac{g_{tt} (X) g_{11} (X)\partial_z X^1 }{ \sqrt{- g_{tt} (X) \left[ g_{zz} (X) + g_{11}(X) ( \partial_z X^1)^2 + g_{22}(X) (\partial_z X^2 )^2 \right ] }}\Big ] &\,=\,& 0 \, , \nonumber \\
\partial_z \Big [  \frac{g_{tt}(X) g_{22} (X) \partial_z X^2 }{ \sqrt{- g_{tt} (X) \left[ g_{zz} (X) + g_{11}(X) ( \partial_z X^1)^2 + g_{22}(X) (\partial_z X^2 )^2 \right ] }}\Big ] &\,=\,& 0 \, . 
\eeqa
These equations are very general and can be solved in any $(X^1,X^2)$ coordinate chart. If we choose the cylindrical chart 
$X^1=r$ and $X^2 = \theta$  we find the simple solutions

i) $ r \,=\, r_0 \,=\, {\rm const} > 0   \quad , \quad \theta \,=\, \theta_0\,=\, {\rm const}$ 

ii) $r \,=\, 0 \quad, \quad g_{\theta \theta}(X) \,=\,0$ \,. 

Here we will consider the second solution corresponding to an infinite static string extended in the $z$ coordinate and localized in the origin of the $(r,\theta)$ plane. The corresponding on-shell action is 
\beq
S^{o.s.}_{NG} =  - \mu \int dt dz \sqrt{-g_{tt} g_{zz}} \, \, \vert_{r = 0} \, .
\eeq
Assuming that $g_{tt}$ and $g_{zz}$ do not depend on $r$ and $\theta$ the Nambu-Goto action takes the form 
\beq
S^{o.s.}_{NG} = - \mu \int dt dz dr d\theta \sqrt{-g_{tt} g_{zz}} \, \frac{\delta(r)}{\pi}
\eeq

A variation in the metric implies the variation 
\beq
\delta S^{o.s.}_{NG} \,=\, \frac12 \int dt \, dz \, dr \, d\theta  \sqrt{-g} \, T^{\mu \nu} \delta g_{\mu \nu}\, ,
\eeq
where 
\beq
T^{\mu \nu} \,=\, - \frac{\mu }{ \sqrt{g_{r r} g_{\theta \theta} }}  \, \frac{ \delta(r)}{\pi}  \begin{pmatrix} (g_{tt})^{-1} & 0 & 0 & 0 \\ 0 & 0 & 0 & 0 \\ 0 & 0 & 0 & 0 \\ 0 & 0 & 0 & (g_{zz})^{-1}  \end{pmatrix} 
\eeq
is the stress energy tensor of the static string.

The variation of the gravity action leads to the Einstein equations of motion 
\beq
R^{\mu \nu} \,-\, \frac{R}{2} g^{\mu \nu} \,-\, \frac{3}{L^2} g^{\mu \nu} \,=\, 8 \pi G_4 \, T^{\mu \nu}
\eeq

\subsection{The conical $AdS_4$ solution}

Assuming that an static rigid string can be described by a static cylindrical symmetric metric, we begin with the most general static cylindrical symmetric metric in AdS background :
\beq
ds^2 = e^{{-2z\over L}} \left [ -e^{2 \nu (r)}dt^2 + e^{2\lambda (r)} dr^2 + e^{2 \psi (r)}d\theta^2  \right ] + e^{2 \lambda (r)} dz^2 \label{metric1} \, ,
\eeq
where $- \infty < z < \infty$. 
This ansatz is motivated by the case of zero cosmological constant \cite{Hiscock:1985uc}. The Einstein equations take the form 
\beqa
&& e^{2{ z \over L}} {e^{-2\,\lambda  }} \left[  \psi''  + \psi'^{2}+ \lambda''  \right] + {3 \over L^2} \left[  e^{-2\,\lambda} - 1 \right] = - 8  G_4 \mu \,  e^{2 \frac{z}{L}} \,  e^{-\lambda}e^{ - \psi} \delta (r) , \label{eqttt} \\ 
&&e^{2{ z \over L}} {e^{-2\,\lambda  }}\left[\nu'  \lambda' + \lambda'  \psi'  + \nu'  \psi'  \right] + {3 \over L^2} \left[  e^{-2\,\lambda} - 1 \right]= 0 \, , \label{eqtrr} \\ 
&& e^{2{ z \over L}} e^{-2\,\lambda  } \left[ \nu'' + \nu'^{2} + \lambda''   \right] + {3 \over L^2} \left[  e^{-2\,\lambda} - 1 \right]= 0 \, , \label{eqtthetatheta}\\
&&e^{2{ z \over L}} e^{-2\,\lambda } \left[- \nu'   \lambda'  - \lambda' \psi' +  \nu''    +   \nu'^{2}  + \nu'  \psi'    + \psi''  + \psi'^{2} \right] + {3 \over L^2} \left[  e^{-2\,\lambda} - 1 \right] \cr
&&= - 8  G_4 \mu \,  e^{2 \frac{z}{L}} \,  e^{-\lambda}e^{ - \psi} \delta (r), \label{eqtzz}\\
&&2\,\frac{\lambda'}{L} = 0 \, \label{eqtrz}.
\eeqa
where $f' \equiv d f / dr$. The equations (\ref{eqtrr}), (\ref{eqtthetatheta}) and (\ref{eqtrz}) imply that $\lambda = 0$ and $\nu$ is a constant that can be chosen zero by a rescaling of the time coordinate. Then equations (\ref{eqttt}) and (\ref{eqtzz}) reduce to the equation
\beq
 \left[  \psi ''  + \psi'^2 \right] e^{\psi} =   \Big [ e^{\psi} \Big ]'' = - 8  G_4 \mu \, \delta (r ) \,, \label{psieq}
\eeq
with solution 
\beqa
e^{\psi} &=& a r + b + {4  G_4 \mu \over  \pi} \int_{-\infty}^{\infty} dk \, {e^{ikr} \over k^2} \cr
&=&  (a - 4 G_4 \mu) r + b
\eeqa
We must have $a=1$ and $b=0$ in order to recover the Anti-de-Sitter space-time in the absence of the string ($\mu =0$). This way we have obtained the metric
\beq
ds^2 = e^{- 2{z \over L}} \left [ -dt^2 + dr^2 + (1-4 G_4 \mu )^2 r^2 d\theta^2 \right ] + dz^2 \, \label{adscone}. 
\eeq
This metric describes a 4-d Anti-de-Sitter space-time with conical deficit as can be seen by redefining the angular coordinate 
\beq
\bar \theta = (1-4 G_4 \mu )   \theta 
\eeq
so that 
\beq
ds^2 = e^{- 2{z \over L}} \left [ -dt^2 + dr^2 + r^2 d \bar \theta^2  \right ] + dz^2  \, ,\label{adscone2}
\eeq
with $0 \leq \bar \theta <  2\pi (1- 4 G_4 \mu)$  corresponding to an angular deficit of $8 \pi G_4 \mu$. A similar angular deficit was obtained before in \cite{Dehghani:2001ft} considering a vortex line configuration in the limit of weak gravitational field. 

In the limit of zero cosmological constant ($L \to \infty$) our solution reduces to the 4-d  metric of a cosmic string  \cite{Hiscock:1985uc} (see also \cite{Vilenkin1994}). Note that the deficit angular must be lower than $2\pi$ so that we have a bound for the string tension :
\beq
0 < \mu < \frac{1}{4 G_4} \, . 
\eeq

\subsection{ The  $2+1$ conical boundary}

Redefining  the transverse radial coordinate 
\beq
r = \frac{\rho^{^{1-4 G_4 \mu}}}{(1-4 G_4 \mu)} 
\eeq
the metric (\ref{adscone}) takes the form 
\beq
ds^2 = e^{- 2{z \over L}} \left [ -dt^2 + \rho^{- 8 G_4 \mu} (d\rho^2 + \rho^2 d\theta^2 )\right ] + dz^2 \, . 
\eeq
Taking the limit $z \to -\infty$ we get the {\it boundary} of 4-d conical AdS space-time : 
\beq
ds^2 =  -dt^2 + \rho^{- 8 G_4 \mu} (d\rho^2 + \rho^2 d\theta^2 ) \, \label{cone} . 
\eeq
This metric can be identified with the conical solution obtained by Desser, Jackiw and 't Hooft \cite{Deser:1983tn} when solving the $2+1$ Einstein equations with zero cosmological constant in the presence of a point particle of mass $M$ localized at the origin.  
This identification implies the relation 
\beq
M = \frac{G_4}{G_3} \, \mu  \,.
\eeq
where $G_3$ is the $2+1$ Newton constant. Note that in the limit of small $G_4 \mu$  we can approximate the metric (\ref{cone}) by
\beq
ds^2 =  -dt^2 + ( 1 - 8 G_4 \mu \ln \rho) \, (d\rho^2 + \rho^2 d\theta^2 ) \, .
\eeq
The logarithmic term is associated with the spatial components of the graviton and is characteristic of $2+1$ fields. 

This way, solving the $3+1$ Einstein equations in the presence of an infinite static string and negative cosmological constant we have found a conical $AdS_4$ space-time whose boundary can be identified with a $2+1$ conical space-time. This result suggest a duality between these spaces similar to the usual $AdS_4/CFT_3$ correspondence. Below we investigate this duality for the case of massive bulk scalar fields in order to see the effect of the conical deficit on boundary scalar operators constructed in the $2+1$ strongly coupled field theory. 

\section{Scalar Green's function in the $2+1$ cone}

In this section we calculate the retarded Green's function of a scalar operator  living in a $2+1$ conical geometry assuming a duality similar to the $AdS_4/CFT_3$ correspondence. Following the AdS/CFT dictionary  a scalar operator  of dimension $\Delta$ living in a $2+1$ conical boundary would be dual to a scalar field with mass $m$ living in the conical $AdS_4$ bulk. The holographic relation between $m$ and $\Delta$ is given by 
\beq
\Delta = \frac32 + \sqrt{ \frac94 + m^2 L^2 } \, , \label{hologrelation}
\eeq
where $L$ is the $AdS_4$ radius. 

Before any calculation it is convenient define the  coordinate $\tilde z = L \, \exp (\frac{z}{L})$ so that the metric (\ref{adscone2}) takes the {\it Poincar\'e} form 
\beq
ds^2 = {L^2 \over \tilde z^2} \left [ -dt^2 + dr^2 + r^2  d\bar \theta^2 + d \tilde z^2 \right ] \, ,\label{adscone3}
\eeq
where $0 < \tilde z < \infty$,  $ 0 \le \bar \theta < \bar \theta_0$ with $\bar \theta_0 = 2 \pi (1-4 G_4\mu )$. The boundary is given by the region  $\tilde z = 0$. 

Consider a massive scalar living in the conical $AdS_4$ geometry with action  
\beq
S= -\kappa \int d^4 x \sqrt{-g} \left [ g^{\mu \nu} \partial_\mu \phi \partial_\nu \phi + m^2 \phi^2 \right]
\eeq
with $\kappa = {1 \over 16 \pi G_4}$. The equation of motion takes the form
\beq
\tilde z^2 \partial_{\tilde z} \left[  \tilde z^{-2}\partial_{\tilde z}\phi \right ] - \partial_t^2 \phi + {1 \over r} \partial_r \left [r \partial_r \phi \right ] + {1 \over r^2} \partial_{\bar \theta}^2 \phi - \frac{m^2 L^2}{\tilde z^2}  \phi =0 \, .
\eeq
The solution for the scalar field regular at $r \to \infty$ can be expanded as 
\beq
\phi (t, r , \bar \theta , \tilde z) = (2 \pi)^{-3/2} \sum_{\lambda} \int d\omega d p \, p  \, e^{-i wt} e^{i \lambda \bar \theta}  J_{\lambda}( p r) {\cal Z}(k ,\tilde z)  \bar \phi_{\lambda}(\omega, p) \, ,\label{generalsolution}
\eeq
where $J_\lambda$ is a Bessel function and ${\cal Z}(k ,\tilde z)$ satisfies the equation
\beq
 \tilde z^2 \partial_{\tilde z}\left [ \tilde z^{-2} \partial_{\tilde z} {\cal Z} (k ,\tilde z) \right ]  +   \left ( k^2 - \frac{m^2 L^2}{\tilde z^2} \right ) {\cal Z} (k ,\tilde z) = 0 \, ,\label{propagatorequation}
\eeq
with $k = \sqrt{\omega^2- p^2}$. Note that we are assuming that the dual operator creates time-like particles and that the scalar field is periodic in $\bar \theta$ even in the presence of an angular deficit, i.e. , 
\beq
\phi(\bar \theta + \bar \theta_0)=\phi(\bar \theta) \quad \rightarrow  \quad 
\lambda = {2n \pi \over \bar \theta_0 } = {n \over (1 - 4 G_4 \mu)} \,. \label{lambdaeq}
\eeq

The solution to eq. (\ref{propagatorequation}) which is real at the boundary and satisfies the {\it incoming wave} condition at the  horizon is 
\beqa
{\cal Z}(k,\tilde z) =  - i D(k)  \tilde z^{3/2} \left[ J_\nu (k\tilde z) \pm i  Y_\nu (k\tilde z) \right] = 
\left\{
\begin{array}{ll} - i D(k) \tilde z^{3/2} H_\nu^{^{(1)}}(k\tilde z) \qquad {\rm for} \quad \omega > 0 \\
- i D(k) \tilde z^{3/2} H_\nu^{^{(2)}}(k\tilde z) \qquad {\rm for} \quad \omega < 0
\end{array} \right. \, ,  \, \, \label{Zsolution}
\eeqa
where $H_\nu^{^{(1)}}$, $H_\nu^{^{(2)}}$ are Hankel functions, $D(k)$ is a real function and  
\beq
\nu = \sqrt{ \frac94 + m^2 L^2 } = \Delta -\frac32 \, .
\eeq

Near the boundary the function ${\cal Z}(k,\tilde z)$ diverges as $\tilde z^{3/2 - \nu}$ so it is convenient to define the boundary at $z=\epsilon$ and make $\epsilon \to 0$ at the very end. The boundary condition for the scalar field can be written as
\beq
\phi (t, r , \bar \theta , \tilde z)\vert_{z = \epsilon} = \epsilon^{3/2-\nu} \phi_0(t,r, \bar \theta) \label{scalarboundcond}
\eeq
where
\beq
\phi_0(t,r, \bar \theta) = (2 \pi)^{-3/2} \sum_{\lambda} \int d\omega d p \, p \, e^{-i wt} e^{i \lambda \bar \theta}  J_{\lambda}( p r) \bar \phi_{\lambda}(\omega, p) \, \label{3dsource}
\eeq
is the boundary field that  couples to a scalar operator ${\cal O}$ with conformal  dimension $\Delta$ satisfying (\ref{hologrelation}). From (\ref{generalsolution}), (\ref{Zsolution}), (\ref{scalarboundcond}) and (\ref{3dsource}) we obtain 
\beq
{\cal Z}(k,\tilde z) =  \frac{\tilde z^{3/2} H_\nu^{^{(1,2)}}(k\tilde z)}{ \epsilon^{3/2} H_\nu^{^{(1,2)}}(k \epsilon)} \,\epsilon^{3/2 - \nu} \,  .
\eeq

The boundary term of the on shell action can be written as 
\beq
S_{Boundary} = -  \kappa \lim_{\epsilon \to 0} \int dt dr  d \bar \theta \sqrt{-g} g^{\tilde z\tilde z} \phi \partial_{\tilde z} \phi |_{\tilde z = \epsilon} \, . \label{onshellaction}
\eeq
Substituting the expansion (\ref{generalsolution}) into this action we find 
\beqa
S_{Boundary} &=& - \frac{\kappa L^2}{2\pi} ( 1 - 4 G_4 \mu ) \,  \sum_{\lambda, \lambda'}  \int_{-\infty}^{\infty} \! \! d\omega \int_{-\infty}^{\infty}\! \!d\omega' \int_0^{\infty} \!dp \, p \int_0^{\infty} \!dp' \,  p' \, \delta_{_{\lambda',-\lambda}} \, \delta (\omega + \omega')  \cr 
&\times&   \, f_{_{ \lambda \, , \,  -\lambda}} (p,p')  \,{\cal F}(\omega, p, \omega',p')  \, \bar \phi_{\lambda}(\omega, p) \bar \phi_{\lambda'}(\omega', p') \, , 
\eeqa
where
\beqa
f_{_{ \lambda \, , \, -\lambda}}(p,p') &=&   \int dr \, r  J_{\lambda}( p r) J_{-\lambda}(p' r)  \, , \label{BesselIntegral}\\
{\cal F}(\omega, p, \omega',p') &=& \lim_{\epsilon \to 0} \left[ \tilde z^{-2} {\cal Z}(k',\tilde z) \partial_{\tilde z} {\cal Z}(k,\tilde z) \right]_{\tilde z=\epsilon}  \, ,
\eeqa
and $k = \sqrt{\omega^2- p^2}$, $k' = \sqrt{\omega'^2- p'^2}$. Using a  real time prescription similar to  \cite{Son:2002sd} we obtain the retarded Green's function 
\beq
G^{R}_{\lambda \lambda'}(\omega, p , \omega', p') =  - \frac{\kappa L^2}{\pi} ( 1 - 4 G_4 \mu ) \, \delta_{_{\lambda',-\lambda}} \, \delta (\omega + \omega')     \, f_{_{\lambda \, , \,  -\lambda}} (p,p')  \, {\cal F}(\omega, p, \omega',p') \, .
\eeq

We can  integrate the Bessel functions in (\ref{BesselIntegral}) finding
\beq
f_{_{\lambda \, , \, -\lambda}} (p,p') =  \cos (\pi \lambda ) \, \frac{\delta(p-p')}{p} - \sin (\pi \lambda) \,  h_\lambda(p,p') \, ,
\eeq
with 
\beq
h_\lambda(p,p') \equiv  -\frac{2}{\pi ( p^2-p'^2)} \left(\frac{p}{p'}\right)^{\lambda}  + \frac{1}{\pi p (p-p')} \frac{\delta(p-p')}{\delta(0)} \label{defh} 
\eeq
On the other hand 
\beqa
{\cal F}(\omega, p , \omega', p') &=& \lim_{\epsilon \to 0} \left [ \epsilon^{1-2\nu} \, \frac{\partial_{\tilde z} \left [ \tilde z^{3/2} H_\nu^{^{(1,2)}}(k \tilde z) \right ] \vert_{\tilde z = \epsilon}}{ \epsilon^{3/2} H_\nu^{^{(1,2)}}(k \epsilon)} \right ]  \cr 
&=& \lim_{\epsilon \to 0} \left [ \left (\frac32- \nu \right ) \epsilon^{-2 \nu} + k \, \epsilon^{1-2\nu} \frac{ H_{\nu-1}^{^{(1,2)}}(k \epsilon)}{H_{\nu}^{^{(1,2)}}(k \epsilon)} \right ] \,.
\eeqa
Using the relation \cite{Abramowitz} 
\beq
H_{\nu}^{^{(1,2)}}(\zeta) = \pm i \csc \pi \nu \left [ - J_{-\nu} (\zeta) + e^{\mp i \pi \nu} J_\nu (\zeta) \right ] \, ,
\eeq
and the Bessel expansions 
\beq
J_{\pm \nu} (\zeta) = \left ( \frac{\zeta}{2} \right )^{\pm \nu} \sum_{\ell=0}^{\infty}
\frac{(-1)^\ell}{\Gamma(\pm \nu + \ell + 1)  \, \ell ! } \left ( \frac{\zeta}{2} \right )^{2 \ell} \, ,
\eeq
we obtain 
\beqa
{\cal F}(\omega, p , \omega', p') = \left\{
\begin{array}{ll} -    (\omega^2-p^2)^{\nu} \, a(\nu) \, \left[ \cos (\pi \nu) - i \sin (\pi \nu) \, \,   {\rm sgn}(\omega) \right ]  + \dots  \quad \nu \notin  \bf Z  \\
       -   (\omega^2-p^2)^{\nu} \, b(\nu) \, \left[ \ln (\omega^2-p^2) - i \pi \, \, {\rm sgn}(\omega) \right]  +  \dots  \qquad \nu \in \bf Z
\end{array} \right. \, ,  \, \,
\eeqa
where 
\beq
a(\nu) =  \frac{2^{1-2\nu} \pi }{\sin (\pi \nu) \Gamma^2(\nu)} \quad , \quad b(\nu) = \frac{2^{1-2\nu}}{\Gamma^2(\nu)} \, ,
\eeq
and the dots indicate divergent terms that are cancelled using holographic renormalization.

In this way we obtain the retarded Green's function 
\beqa
G^R_{\lambda \lambda'}(\omega, p , \omega', p') &=&   \frac{\kappa L^2}{\pi} a(\nu) ( 1 - 4 G_4 \mu ) \, \delta_{_{ \lambda',-\lambda}} \, \delta (\omega + \omega') \left [ \cos (\pi \lambda ) \, \frac{\delta(p-p')}{p} 
- \sin (\pi \lambda) \,  h_\lambda(p,p') \right ] \cr 
&\times&  (\omega^2-p^2)^{\nu} \left[ \cos (\pi \nu) - i \sin (\pi \nu) \, \,   {\rm sgn}(\omega) \right ] \label{resultGR1}
\eeqa
for $\nu \notin  \bf Z$ and
\beqa
G^R_{\lambda \lambda'}(\omega, p , \omega', p') &=&  \frac{\kappa L^2}{\pi} b(\nu) ( 1 - 4 G_4 \mu ) \, \delta_{_{\lambda',-\lambda}} \, \delta (\omega + \omega') \, \left [ \cos (\pi \lambda ) \, \frac{\delta(p-p')}{p} 
- \sin (\pi \lambda) \,  h_\lambda(p,p') \right ] \cr 
&\times&  (\omega^2-p^2)^{\nu} \left[ \ln (\omega^2-p^2) - i \pi \, \, {\rm sgn}(\omega) \right] \,  \label{resultGR2}
\eeqa
for $\nu \in \bf Z$, where $h_\lambda(p,p')$ is given in eq. (\ref{defh}) and $\nu = \Delta - 3/2$. 

These results are valid as long as the time-like condition $\omega^2 > p^2$ is satisfied. In the space-like case ($\omega^2 < p^2$) the imaginary terms in (\ref{resultGR1}) and (\ref{resultGR2}) disappear while the real terms can be obtained inverting the sign of $\omega^2-p^2$ and substituting $\cos(\pi \nu)$ by $1$. 

The conical deficit, represented by $G_4 \mu$ , not only modifies the global factor of the retarded Green's function but also introduces a term proportional to $\sin (\pi \lambda)$ with $\lambda= n/(1-4G_4\mu)$ that not only breaks conformal invariance but makes the retarded Green's function now depend on both momenta $p$ and $p'$. 
We will see that even in the absence of this term any Green's function in the 3-d conical space-time would depend separately on the space vectors $\vec{x} = (r \cos \theta , r \sin \theta) \,, \, \vec{x}' = (r'\cos \theta' , r' \sin \theta')$ and not only on the difference $\vec{x} - \vec{x}'$.  


\section{Conical Green's functions in space coordinates}

As we mentioned before, the conical deficit breaks not only  conformal symmetry but also translation invariance. Below we will investigate how this breaking occurs. For simplicity we will consider discrete values for the string tension 
\beq
\mu = \frac{1}{4G_4} \left (  1 - \frac{1}{k}  \right ) \qquad \qquad  k = 1,2,\dots \, , 
\eeq
so that we have integer values for $\lambda$ and $\lambda'$. Note that the term proportional to $\sin (\pi \lambda)$ vanishes. Now we need to know how  the the Green's function depend on  the space coordinates $r$ and $\bar \theta$. For this purpose we perform the  Bessel-Fourier transformation 
\beqa 
\delta(\omega + \omega') G^R(\omega,  r , \bar \theta ,   r',   \bar \theta') &=& \frac{1}{(2 \pi)^2} \sum_\lambda \sum_{\lambda'} \, e^{i\lambda \bar \theta} e^{i \lambda' \bar \theta'}  \int dp \, p \int dp' \,  p'  \cr 
&\times& J_\lambda(pr) J_{\lambda'}(p'r') \, G^R_{\lambda \lambda'}(\omega, p , \omega', p') \, . 
\eeqa
The green function  takes the form 
\beqa
G^R(\omega,  r , \bar \theta ,   r',   \bar \theta') = \left\{
\begin{array}{ll} \frac{\kappa L^2  }{ 4 \pi^3 k } a(\nu)  \left[ \cos (\pi \nu)  - i \sin (\pi \nu)  {\rm sgn} (\omega) \, \right] \Omega ( k , \nu , \omega,  r , r', \bar \theta -  \bar \theta')  \quad \nu \notin  \bf Z  \\
       \frac{\kappa L^2  }{ 4 \pi^3 k } b(\nu) \left[ \partial_\nu - i \pi \, {\rm sgn} (\omega) \, \right] \Omega ( k , \nu , \omega,  r , r', \bar \theta -  \bar \theta')  \qquad \nu \in \bf Z
\end{array} \right. \, ,  \, \, \label{retGreenomega}
\eeqa
where 
\beq
\Omega ( k , \nu , \omega,  r , r', \bar \theta -  \bar \theta') = \sum_n e^{i k n (\bar \theta - \bar \theta')}  \int dp \, p \, J_{k n }(pr) J_{k n}(pr') (\omega^2-p^2)^{\nu} \, .
\eeq

Using the representation 
\beq
J_{kn} (pr) J_{kn}(pr') = \frac{1}{2\pi} \int_0^{2\pi} d \alpha \cos (k n \alpha) J_0 
\left ( p \sqrt{r^2 + r'^2 - 2 r r' \cos \alpha} \right )
\eeq
and the integral 
\beq
\int_0^\infty dp \, p \, J_0(p a) (\omega^2 - p^2)^\nu = \frac{2^{\nu+1} }{\pi} \Gamma(\nu + 1) \sin (\pi \nu) (i |\omega| )^{\nu+1} a^{-\nu -1} K_{\nu + 1}(- i \omega a), 
\eeq
we obtain 
\beqa
 \Omega ( k , \nu , \omega,  r , r', \bar \theta -  \bar \theta') &=& \frac{2^{\nu}}{  \pi^2} \Gamma(\nu + 1) \sin (\pi \nu) (i |\omega| )^{\nu+1}\int_0^{2\pi} d \alpha \sum_n \cos [ k n (\bar \theta - \bar \theta')] \cos (k n \alpha )   \cr 
&\times&  \left ( r^2 + r'^2 - 2 r r' \cos \alpha \right )^{-\frac{\nu+1}{2}} K_{\nu+1} ( - i \omega \sqrt{r^2 + r'^2 - 2 r r' \cos \alpha} ) \,, \label{Omegaeq}
\eeqa
where $K_\nu(w)$ is the modified Bessel function. Using this result in (\ref{retGreenomega}) we find the retarded Green's function
\beqa
G^R ( \omega,  r , r', \theta -  \theta') &=& \frac{\kappa L^2  }{ 4 \pi^5 k }  \frac{2^{1 - \nu} \nu }{  \Gamma(\nu) } \left[ \cos (\pi \nu) - i \pi \sin (\pi \nu) {\rm sgn} (\omega) \right] (i |\omega| )^{\nu+1} \cr 
&\times& \int_0^{2\pi} d \alpha \left [ \sum_n \cos [ k n (\bar \theta - \bar \theta')] \cos (k n \alpha ) \right ]  \cr 
&\times&  \left ( r^2 + r'^2 - 2 r r' \cos \alpha \right )^{-\frac{\nu+1}{2}} K_{\nu+1} ( - i \omega \sqrt{r^2 + r'^2 - 2 r r' \cos \alpha} ) \,, \label{retardedalpha}
\eeqa
expressed as an integral of the extra angular coordinate $\alpha$. This result is valid for integer and non-integer $\nu$. In the absence of deficit ($k=1$) we can use the completeness relation 
\beq
\sum_n  \cos ( n \alpha ) \cos (n \beta ) = 2 \pi \delta ( \alpha - \beta ) 
\eeq
to obtain
\beqa
G^R_{(k=1)} ( \omega,  r , r', \theta -  \theta') &=& \frac{\kappa L^2  }{ 2 \pi^4 k }  \frac{2^{1 - \nu} \nu }{  \Gamma(\nu) } \left[ \cos (\pi \nu) - i \pi \sin (\pi \nu) {\rm sgn} (\omega) \right] \cr 
&\times& (i |\omega| )^{\nu+1}  |\vec{r} - \vec{r}'|^{-\nu -1} K_{\nu+1} ( - i \omega | \vec{r} - \vec{r}' | ) \,,
\eeqa
where $| \vec{x} - \vec{x}' |= \sqrt{r^2 + r'^2 - 2 r r' \cos (\theta - \theta')}$ is the distance between two points in the 2-d cone. After a Fourier transform in the $\omega$ coordinate takes the form 
\beq
G^R_{k=1}(t -t', r , r', \theta -  \theta'  ) \sim \left ( (t-t')^2 - | \vec{r} - \vec{r}' |^2 \right )^{- \Delta} \label{localresult}
\eeq
which is the expected result for a $2+1$ conformal field theory \cite{D'Hoker:2002aw}. 

In the presence of a conical deficit ($k > 1$) it is very difficult to solve the integral in $\alpha$ at any energy $\omega$. However, this can be done in the low energy regime ($\omega \ll 1$) and for integer $\nu$. Using the approximation
\beq
K_{\nu + 1}(\zeta \to 0  ) \approx  2^\nu \Gamma(\nu + 1 ) \zeta^{- \nu -1} \, ,
\eeq
eq. (\ref{retardedalpha}) takes the form 
\beqa
G^R (\omega \to 0) &=& - \frac{\kappa L^2  }{ 2 \pi^5 k }   \nu^2 \, (r'^2)^{- \nu - 1} \int_0^{2\pi} d \alpha \, \frac{\left [ 1 + 2 \sum_{n=1}^{\infty} \cos [ k n (\bar \theta - \bar \theta') ] \cos (k n \alpha ) \right]}{ \left [ 1  - 2 \xi  \cos \alpha  + \xi^2  \right ]^{\nu+1}} \cr
&=& - \frac{\kappa L^2  }{  \pi^4 k }  \nu^2  (r'^2)^{- \nu - 1} \frac{\xi^{2\nu}}{(1-\xi^2)^{2\nu+1}} \sum_{i=0}^\nu \frac{(2\nu-i)!}{i! \, \nu ! \, (\nu - i)!} \left ( \frac{1-\xi^2}{\xi^2} \right )^i \xi^{i-\nu} \cr 
&\times& \left( \frac{\partial}{\partial \xi} \right)^i 
\left[  \frac{\xi^\nu (1 - \xi^{2k})}{1 - 2 \xi^k \cos [k (\bar \theta - \bar \theta')] + \xi^{2k}}  \right]  \, ,  
\eeqa
where $\xi=r/r'$ and we used the integral 
\beq
\int_0^{2\pi} d \alpha \, \frac{\cos (m \alpha)}{ ( 1 - 2 \xi \cos \alpha + \xi^2 )^{\nu + 1}} = 2 \pi \frac{\xi^{2 \nu + m}}{(1-\xi^2)^{2\nu+1}} \sum_{i=0}^\nu \frac{(\nu + m)! (2 \nu - i )! }{i! (\nu + m - i )! \nu ! (\nu  - i)!} \left ( \frac{1 - \xi^2}{\xi^2} \right)^i
\eeq
valid for $\xi^2 \le 1$, $m \ge 0$ and the series 
\beq
\sum_{n=1}^{\infty} \, \cos(n x) y^n = \frac12 \left ( \frac{1-y^2}{1-2y\cos x + y^2} - 1 \right )  
\eeq
valid for $y^2 \le 1$ \cite{Gradshteyn}. It is very interesting to analyze the case $\nu=1$ (corresponding to $\Delta=5/2$) that gives 
\beqa
G^R(\omega \to 0) \vert_{\nu=1} &=& - \frac{\kappa L^2  }{  \pi^4 k } (r'^2-r^2)^{-3} \left [ r'^{2k} - 2 r'^k r^k \cos [k (\bar \theta - \bar \theta')] + r^{2k} \right ]^{-2}        \cr 
&\times& \Big \{ (r'^2 + r^2)(r'^{4k}-r^{4k}) - 4k(r'^2-r^2)r'^{2k}r^{2k} \cr 
&+& [ 2(k-1)r'^2 - 2(k+1)r^2 ] r'^{3k}r^k  \cos [k (\bar \theta - \bar \theta')] \cr 
&+& [ 2(k+1)r'^2 - 2(k-1)r^2 ] r'^k r^{3k}  \cos [k (\bar \theta - \bar \theta')]   \Big \} \,.
\eeqa

This Green's function  (for $k>1$)  depends separately on the variables $r$, $r'$, $(\bar \theta-\bar \theta')$ and not only on the distance $| \vec{x} - \vec{x}' |= \sqrt{r^2 + r'^2 - 2 r r' \cos (\theta - \theta')}$.  This means that  a  scalar perturbation on a point $\vec{x} = (r \cos \theta , r \sin \theta)$ produce an effect on a point $\vec{x}' = (r'\cos \theta' , r' \sin \theta')$ that depends not only on the distance between them  but also on the position of the perturbation.  This is a consequence of the breaking of translation invariance.  A similar result can be obtained for other values of $\nu$. In the $k=1$ case (corresponding to $\mu = 0$) we recover the result (\ref{localresult}). 

Note that in the limit $\theta \to \theta'$, $r \to  r'$ the Green's function reduces to 
\beq
G^R (\omega \to 0) \vert_{\nu=1} \approx - \frac{\kappa L^2  }{  \pi^4 k^2 }(r'-r)^{-4} = \frac{1}{k^2} G^R_{k=1}(\omega \to 0) \vert_{\nu=1} \,.
\eeq
which means that the conical Green's function  ($k > 1$) diverges in the same way as the flat  Green's function ($k=1$). The only difference between the singularities is a factor of $1/k^2$. This result is very different from that obtained solving the Green's function of a $2+1$ massless scalar field \cite{Souradeep:1992ia} that would correspond to a case $\nu=-1$ which is forbidden in the duality. The Green's function obtained in \cite{Souradeep:1992ia} has a logarithmic dependence so that the singularity can be removed. In our case the singularity can not be removed because of the power dependence of the Green's functions of operators with dimension $\nu > 0$.

\section{Conclusions}

In this work we found a conical $AdS_4$ spacetime by solving the 4-d Einstein equations in the presence of an infinite string and negative cosmological constant. We identified its boundary with the 3-d conical spacetime found by Deser, Jackiw and 't Hooft. Our results suggest a correspondence between fields living in a conical $AdS_4$ spacetime and operators living in a $2+1$ cone. We used this correspondence to calculate the retarded Green's function of scalar operators in the 3-d conical spacetime. 

It is important to remark that the duality proposed in this paper is phenomenological in the sense that the conical $AdS_4$ solution is not derived from string theory as the near-horizon geometry of some brane set-up. For this reason we still don't know the particular details of the bulk and boundary theories. Nevertheless, we believe that it is worth to explore this non-conformal duality since the conical space is natural solution of 3-d gravity and it can be useful to investigate structural defects in condensed matter systems (like graphene). 
For this purpose it  would be interesting to explore further this duality for the case  of spin $1$ and $2$ operators and look for possible applications in $2+1$ strongly coupled field theories.


\bigskip

\noindent {\bf Acknowledgments:} The authors are financially supported by CNPq.

\end{document}